\documentclass[aps,prl,twocolumn,groupedaddress,dvips]{revtex4}
\usepackage[dvips]{graphicx}
\begin{document}
\title{Suppression of Spin-Orbit Scattering in Strong-Disordered Gold Nanojunctions}
\author{A. Anaya, M. Bowman, and D. Davidovi\'c}
\affiliation{Georgia Institute of Technology, Atlanta, GA 30332}
\date{\today}
\begin{abstract}
We discovered that spin-orbit scattering in strong-disordered gold
nanojunctions is strongly suppressed relative to that in
weak-disordered gold thin films. This property is unusual because
in weak-disordered films, spin-orbit scattering increases with
disorder. Granularity and freezing of spin-orbit scattering inside
the grains explains the suppression of spin-orbit scattering. We
propose a generalized Elliot-Yafet relation that applies to
strong-disordered granular regime.
\end{abstract}
\maketitle

The field of spintronics has recently emerged as a potential
alternative to conventional charge-based electronics.~\cite{wolf1}
What sets spintronics apart is the explicit study or use of the
electron spin degree of freedom.  A challenge in spintronics is
the finite lifetime of spin-polarized current, since electron
spins can flip in normal metals and semiconductors.

It is generally accepted that a spin-orbit (SO) interaction,
through the so called Elliot-Yafet mechanism,~\cite{elliot,yafet}
causes spin-flip scattering in weak-disordered metals. In this
mechanism, the SO scattering time ($\tau_{so}^{ey}$) is
proportional to the momentum relaxation time $\tau$,
$\tau_{so}^{ey}=\tau/\alpha$, which is known as the Elliot-Yafet
relation. The scattering ratio $\alpha \ll 1$ represents the
spin-flip probability during the momentum relaxation time. It
depends on the atomic number, band structure, and to a lesser
extent, on sample preparation techniques. It has recently been
demonstrated that the Elliot-Yafet relation agrees with measured
SO scattering time in a wide range of weak-disordered metallic
samples.~\cite{jedema3}

In this paper we investigate SO scattering in strong-disordered
metals, that is, in metals where conduction electrons undergo
transition into Anderson localized states at low temperatures. We
find that the relation between disorder and SO scattering time in
the strong-disordered regime is qualitatively different from that
in the weak-disordered regime. We observe a strong enhancement of
the SO scattering time compared to that in weak-disordered
samples.
We propose that the enhancement of the SO scattering time
arises from granularity, as follows.

Consider a 3D granular system composed of grains with average
diameter $D$ and average grain-to grain resistance $R_g$. If $R_g$
is larger than $R_Q=h/e^2=25.8k\Omega$, within a factor of order
one, then the system is strong-disordered. If $R_g<R_Q$, within a
factor of order one, then the system is
weak-disordered.~\cite{beloborodov1}

By definition, $R_g$ is larger than the resistance inside the
grains. The dwell time of an electron on any given grain ($t_D$)
is roughly $t_D=t_H R_g/R_Q$, where $t_H=h/\delta$ is the
Heisenberg time ($\delta$ is the level spacing).

\begin{figure}
\includegraphics[width=0.47\textwidth]{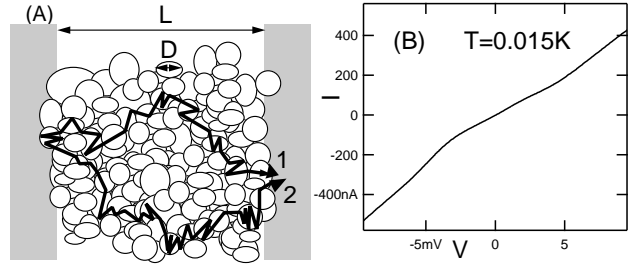}
\caption{A. Sketch of a granular sample and two semi-classical
electron trajectories traversing the sample. B. I-V curve of a
sample smaller than the localization length.~\label{fig1}}
\end{figure}

We consider small strong-disordered granular samples, in which the
electron localization length is larger than sample size. In these
samples, electrons at the Fermi level are spatially extended
through the sample.

 We discuss SO scattering time of
electrons at the Fermi level and at zero temperature. We assume
the grains are ballistic so that momentum relaxation is dominated
by surface scattering. So the Elliot-Yafet relation predicts a SO
scattering time of $\tau_{so}^{ey}=D/(\alpha v_F)$. We argue that
$\tau_{so}^{ey}$ is not a good estimate of the spin-orbit
scattering time if the grains are sufficiently small and $R_g$ is
large enough.

If the grains were completely isolated,  the strength of SO
interaction would be governed by a dimensionless parameter
$t_H/\tau_{so}^{ey}$, and SO scattering inside the grains would be
weak if $t_H<\tau_{so}^{ey}$.~\cite{brouwer,matveev}  In this
case, the electronic wavefunctions of the grain are nearly all
spin-up or all spin-down.

Reducing the grain diameter decreases both $t_H$ and
$\tau_{so}^{ey}$, as $t_H\sim \frac{D^3}{v_F\lambda_F^2}$ and
$\tau_{so}^{ey}\sim\frac{D}{\alpha v_F}$, respectively. Since
$t_H$ decreases faster than $\tau_{so}^{ey}$, a borderline
diameter $D^\star$ exists, below which SO scattering is weak. From
$t_H\sim\tau_{so}^{ey}$, we obtain $D^\star
=\lambda_F/\sqrt{\alpha}$.

In a granular system as in Fig.~\ref{fig1}-A, once an electron is
localized within any given grain, its motion is governed by the
wavefunctions of the grain. If $D<D^\star$, the spin-flip
probability inside the grain would be small, even if
$t_D>\tau_{so}^{ey}$, because SO scattering in individual grains
freezes after the Heisenberg time.~\cite{brouwer,matveev}

Since SO scattering inside the grains with $D< D^\star$ is weak,
we propose that  electrons at the Fermi level at zero temperature
may flip spin only when they hop between neighboring grains, with
a spin-flip probability $\alpha$. This leads to an estimate

\begin{equation}
\tau_{so}=\frac{t_D}{\alpha}\sim \frac{D^3}{\alpha v_F
\lambda_F^2}\frac{R_g}{R_Q}. \label{EYR}
\end{equation}

This equation generalizes the Elliot-Yafet relation for granular
systems. An interesting feature of this equation is that the SO
scattering time is proportional to the resistance between grains.
The resistivity is $\rho\sim R_gD$, hence an increase in
resistivity leads to an increase in SO scattering time, a behavior
opposite to that found in weak-disordered homogeneous metals,
since $\tau_{so}$ is enhanced at the expense of dwell time.

\begin{figure}
\includegraphics[width=0.45\textwidth]{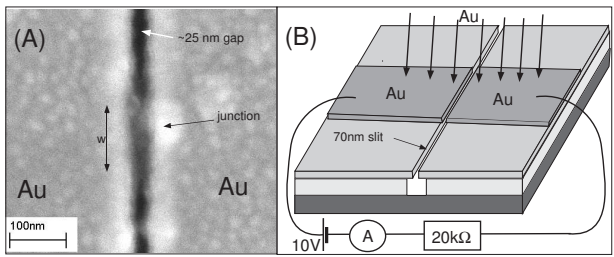}
\caption{A. Image of a strong-disordered Au nano-junction. B.
Fabrication of strong-disordered Au nanojunctions.~\label{fig2}}
\end{figure}

We use electron transport in strong-disordered gold nanojunctions
to investigate SO scattering in the strong-disordered regime. An
image of one nanojunction, from a scanning electron microscope, is
shown in Fig.~\ref{fig2}-A. We create these nanojunctions by
making electric contacts between two Au films at large bias
voltage.~\cite{anaya,bowman}

To summarize, Au atoms are deposited in high vacuum over two bulk
Au films separated by a $\sim 70$nm slit, as sketched in
Fig.~\ref{fig2}-B. The applied voltage is 10 Volt and the current
is measured during the deposition, to detect the moment of
contact, at which point the evaporation is stopped and the voltage
is reduced. Large voltage introduces strong-disorder in the
nanojunction, through processes such as electromigration, surface
atom diffusion, and intermixing with $H_2O$ and $O_2$
molecules.~\cite{anaya}

Au films to the left and right of the nanojunction are good metals
with resistivity $\approx 35\mu \Omega cm $. Through scanning
electron microscopy combined with in situ transport measurements,
we determined that the nanojunctions were homogeneous at length
scale comparable to the gap size (we determined that the sample
resistance was inversely proportional to width w in
Fig.~\ref{fig1}). The resistivity of the material inside the
nanojunction was estimated to be $\rho\approx 1\cdot 10^5\mu\Omega
cm$.~\cite{bowman} This value is larger than the so called
"maximum metallic resistivity" of $200\mu\Omega cm$. Thus, the
nominal transport mean free path ($l$), obtained from
$\rho=mv_F/ne^2l$, is $l\approx 0.01\AA$, much shorter than the
Fermi wavelength.

The short mean free path indicates that Au in the nanojunction is
very disordered. We have explained the disorder by granularity and
the grain size much smaller than the nanojunction
dimensions.~\cite{bowman} In particular, the disorder could not be
amorphous as Au did not alloy with the impurities that were
present in sample fabrication ($H_2O$ or $O_2$). Our imaging
resolution, however, was insufficient to determine the grain
diameter. Recently, it has been demonstrated that large bias
voltages applied to Au surfaces can induce electrochemical
processes that lead to formation of Au nanoparticles of diameter
on the order of a few nanometer,~\cite{nieto,grose} and these
nanoparticles self-assemble into a granular
structure.~\cite{grose}

In the prior work,~\cite{bowman} we showed that the electron
localization length in the nanojunction is finite at low
temperatures. To summarize, samples with resistance larger than
approximately $R_Q$ displayed Coulomb blockade. The width of these
samples was smaller than approximately 50 nm. The Coulomb blockade
was attributed to electron tunnelling on and off the localized
electronic wavefunctions inside the junction. From the temperature
dependence of the conductance in the Coulomb blockade, we
determined that the localization length exceeded 20nm.

In addition, samples with resistance smaller than about $0.5R_Q$
did not exhibit Coulomb Blockade at $T=0.015K$. The I-V curve of
these samples at $T=0.015K$ was linear around zero bias voltage.
The dependence of the Coulomb blockade on sample resistance was
analogous to that found in strong-disordered $InO_X$ mesoscopic
semiconductors.~\cite{chandrasekhar}

The absence of Coulomb blockade in low resistance samples was
explained by the localization length exceeding the nanojunction
length. In this regime, the electronic wavefunctions extended from
one reservoir to another reservoir and quantum electron transport
could be described using models based on weak-disorder theories.
For this letter, we select these low resistance samples, because
quantum interference effects are easier to interpret in these
samples.

Note that  in a granular system, the transport mean free path can
be much shorter than the Fermi-wavelength even in the metallic
state at zero temperature.~\cite{Imry} Thus, it is not unphysical
that the electron wavefunctions extend through the nanojunction,
despite the fact that the transport mean free path is
$\ll\lambda_F$.

The standard technique to measure $\tau_{SO}$ in disordered
conductors is weak-localization,~\cite{washburn} the effect that
originates from the orbital effect of the applied magnetic field
on the electron wavefunctions. In our samples, however, the
nominal transport mean free path is shorter than $\lambda_F$, and
we show that in this case the Zeeman effect is stronger than the
orbital effect. We introduce a new technique to measure a
$\tau_{so}$ lower bound.

We determined that $\tau_{so}$ lower bound varies weakly among low
resistance samples, down to the samples with resistance below
$0.1R_Q$. The I-V curve of sample 2 (Fig.~\ref{fig1}-B) is linear
at zero bias voltage. It has only a weak suppression of
differential conductance near zero bias voltage, which is a
Coulomb-Blockade precursor.~\cite{anaya}

\begin{figure}
\includegraphics[width=0.32\textwidth]{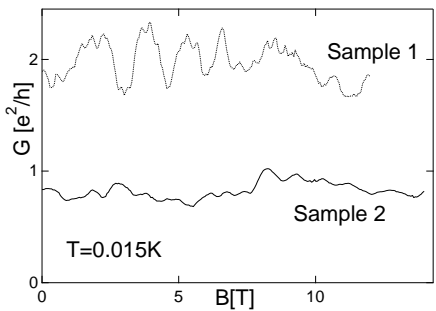}
\caption{A: Differential conductance versus magnetic
field.~\label{fig3}}
\end{figure}

A lock-in technique, with a $2\mu V$ excitation voltage, measures
the differential conductance ($G$). Fig.~\ref{fig3}-A shows $G$
versus $B$ in samples 1 and 2. The conductance clearly exhibits
fluctuations with the magnetic field. The fluctuations are
consistent with universal conductance fluctuations or what are
known as magnetofingerprints, because: 1) the amplitude of the
fluctuations is $\sim e^2/h$; 2) fluctuations are reproducible
with field sweep; and 3) fluctuations are uncorrelated when
samples are thermally cycled.

\begin{figure}
\includegraphics[width=0.47\textwidth]{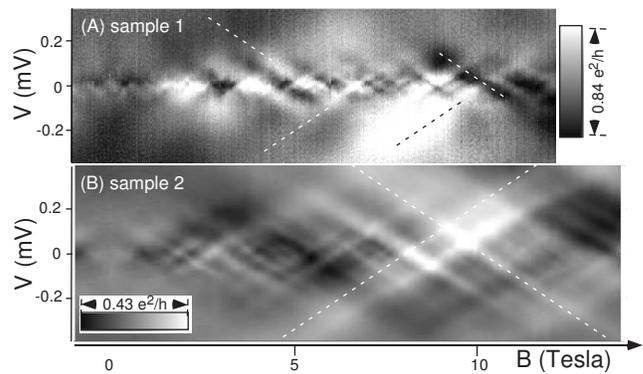}
\caption{Differential conductance versus magnetic field and bias
voltage in samples 1 and 2 at $T=0.015K$. ~\label{fig4}}
\end{figure}

Magnetofingerprints in our samples differ from those in
weak-disordered metals, in that they are caused by the Zeeman
effect, not by the Aharonov-Bohm effect. The gray-scale image in
Fig.~\ref{fig4} displays differential conductance versus bias
voltage and magnetic field ($G(V,B)$). The main signature of the
data is a structure in conductance which shifts linearly with $V$
and $B$, with pronounced lines in $V-B$ parameter space. Some of
the lines are highlighted with dashed lines of the form $eV\pm
2\mu_B B=const$. By comparison, in weak disordered metallic
samples where the mean free path is longer than Fermi wavelength,
the fluctuations with field and the fluctuations with voltage are
uncorrelated.~\cite{washburn}

Consider quantum interference among two semiclassical electron
pathways through the sample, depicted in Fig.~\ref{fig1}-A. The
interference depends on the phase difference $\phi$ between the
probability amplitudes of the trajectories, and can be
constructive or destructive, depending on whether $\phi=2n\pi$ or
$\phi=(2n+1)\pi$,  where $n=0,\pm1,...$. For reference, we recall
that the characteristic magnetic field for the Aharonov-Bohm
effect is given by the field for a flux quantum $\Phi_0 =h/2e$
over the sample area, $B_{AB}=\Phi_0/L^2$.~\cite{washburn}

In a magnetic field, Zeeman splitting causes spin-up and spin-down
electrons to have different Fermi wavelengths, hence
spin-dependent contribution to $\phi$. We find the contribution is
$ \sigma \mu_B B(t_2-t_1)/\hbar$, where $\sigma=\pm 1$ corresponds
to the spin direction and $t_{1,2}$ are the times of flight along
the trajectories. Typical times of flight are $L^2/v_F l$, where
$l$ is the transport mean free-path. The Zeeman effect becomes
significant when $\phi\sim 1$, and the characteristic field for
the Zeeman effect is $B_Z=\frac{l}{\lambda_F}\frac{\Phi_0}{L^2}$.
Thus, conductance fluctuations are spin-based if
$B_Z/B_{AB}=l/\lambda_F<1$. Our sample parameters are such that
$B_Z/B_{AB}=l/\lambda_F<1$.

If the bias voltage is nonzero, then electrons injected from the
Fermi level have a voltage dependent contribution to $\phi$, which
is $\int_0^{t_1} eV(t)dt/\hbar-\int_0^{t_2} eV(t)dt/\hbar$. In
Ohmic samples, the voltage drop is linear in space, and the
contribution becomes $eV(t_1-t_2)/(2\hbar)$. Thus,
\[\phi=eV(t_1-t_2)/(2\hbar)+\sigma\mu_B B(t_1-t_2)/\hbar .\] For any
given pair of trajectories, the voltage dependent contribution is
proportional to the spin-dependent contribution. It follows that
if both $V$ and $B$ are varied with a constraint that $eV+\sigma
2\mu_B B=const$, then the interference of an electron with spin
$\sigma$ is unchanged. As a result, the conductance of electrons
with spin $\sigma$ is constant when $eV+\sigma 2\mu_B B=const$,
explaining Fig.~\ref{fig4}.

We have neglected SO scattering in the analysis. In the following
paragraphs, we take SO scattering into account. SO scattering does
not destroy phase coherence~\cite{altshuler} and we need to obtain
$\phi$ in the presence of SO scattering.

Assume SO scattering to be strong, $\tau_{so}\ll t$, where $t$ is
the typical time of flight defined above. In this case, phase
coherence survives only in the singlet channel, in which an
electron traversing one trajectory with spin-up interferes with
itself after traversing a second trajectory with
spin-down.~\cite{altshuler} The phase-shift in the singlet channel
is
\[\phi_{SO}=eV(t_1-t_2)/(2\hbar)+\sigma\mu_B
B(t_1+t_2)/\hbar .\]

Hence, the voltage-dependent contribution to the phase is not
proportional to the field-dependent contribution. The ratio of
these two contributions varies randomly among different pairs of
trajectories. So
 fluctuations in conductance versus field should
be uncorrelated with fluctuations in conductance versus voltage.
However, Fig.~\ref{fig4} is contrary to what one would expect for
strong SO scattering ($\tau_{so}\ll t$). Then it follows that the
SO scattering is not strong ($\tau_{so}>t$).


The typical time of flight ($t$) can be obtained from the
correlation energy $E_C$ as $E_C=h/t$. $E_C$ is the interval of
electronic energies within which the electronic wavefunctions are
correlated in space. It can be measured from the correlation
voltage ($V_C$) as $E_C=V_C/e$,~\cite{washburn}  the
characteristic voltage scale for the fluctuations in conductance
with bias voltage. $V_C$ is roughly equal to the spacing between
lines in Fig.~\ref{fig4} along the direction parallel to the
bias-voltage axes.


The correlation voltage is obtained from the voltage correlation
function $Y(V)$, as $Y(V_C)=0.5Y(0)$, where
$Y(V)=\overline{G(V,B')G(0,B')}-\overline{
G(V,B')}\,\,\overline{G(0,B')}$ and averaging is over $B'$. We
obtain  $V_C(0)=31\mu V$ and $104\mu V$ in samples 1 and 2,
respectively. It follows that $\tau_{so}>1.3\cdot 10^{-10}$s and
$\tau_{so}>4\cdot 10^{-11}$s in samples 1 and 2, respectively.
 By
comparison, $\tau_{so}$ measured in weak-disordered gold films
with resistivity $\rho\approx 66\mu\Omega cm$ is $1.9\cdot
10^{-13}$s,~\cite{bergman1} at least three orders of magnitude
shorter than $\tau_{so}$ in our nanojunctions. This is the main
finding of this paper.

The suppression of SO scattering in our samples can be explained
by the granular model described in the introduction. We extract
$\alpha$ from $\tau_{so}$ measured by weak-localization in thin
films of Au,~\cite{bergman1} $\alpha\approx 5\cdot 10^{-3}$. In
this case we obtain $D^\star\sim\lambda_F/\sqrt{\alpha}\approx
7nm$. The scattering ratio $\alpha$ can also be extracted from
energy level spectroscopy of nanometer scale Au
grains.~\cite{petta} In this case, we find that $\alpha$ varies
among different grains, $0.01<\alpha<0.05$. Thus
$2.2$nm$<\lambda_F/\sqrt{\alpha} <5$nm.


For example, assume $D=3nm$ and $\alpha=0.01$. From resistivity we
obtain $R_g\approx\rho/D =4\cdot 10^5\Omega$, and Eq.~\ref{EYR}
predicts $\tau_{so}\approx 2.3\cdot 10^{-10}$s.




In conclusion, we have demonstrated that the magnetofingerprints
in strong-disordered Au nanojunctions are spin-based. The
signature of spin-based magnetofingerprints is a structure in
conductance that shifts linearly with bias voltage and magnetic
field. The linear structure requires that the transport time (or
the dephasing time) be shorter than or comparable to the SO
scattering time, which we use to estimate a lower bound of the SO
scattering time. The SO scattering time in strong-disordered
samples is enhanced by at least three orders of magnitude relative
to that in weak-disordered thin films. Granularity and frozen
spin-orbit scattering inside the grains suppresses spin-orbit
scattering. We propose a generalization of the Elliot-Yafet
relation that applies to strong-disordered granular samples and
that agrees with our observations.

As a final note, electron spins confined in quantum dots have been
proposed as candidate quantum bits, because spin is stable for
zero-dimensional systems.~\cite{khaetskii} In this paper, we show
that, without quantum dots, a distributed strong-disordered system
has a similar spin stability. This may be an alternate route for
the fabrication of solid-state devices with high spin-stability.

We thank A. L. Korotkov for useful discussions. This work was
performed in part at the Cornell Nanofabrication Facility, (a
member of the National Nanofabrication Users Network), which is
supported by the NSF, under grant ECS-9731293, Cornell University
and Industrial affiliates, and the Georgia-Tech electron
microscopy facility. This research is supported by the David and
Lucile Packard Foundation grant 2000-13874 and the NSF grant
DMR-0102960.

\bibliography{sc}

\end{document}